\documentclass[pra,aps,twocolumn,showpacs]{revtex4-1}
\usepackage{amsmath,amssymb,graphicx}
\usepackage{epsfig}
\usepackage{textcomp}
\usepackage{color}
\usepackage{epstopdf}
\usepackage{hyperref}

\hypersetup{pdfpagemode=UseNone,colorlinks=true,linktoc=page,pdfborder={0 0 0},linkcolor=blue,citecolor=blue}

\begin{document}

\title{Quantum versus classical effects in the chirped-drive discrete
nonlinear Schrodinger equation}
\author{Tsafrir Armon and Lazar Friedland}
\email{lazar@mail.huji.ac.il}
\affiliation{Racah Institute of Physics, Hebrew University of Jerusalem,
Jerusalem 91904, Israel}
\begin{abstract}
A chirped parametrically driven discrete nonlinear Schrodinger equation is
discussed. It is shown that the system allows two resonant excitation
mechanisms, i.e., successive two-level transitions (ladder climbing) or a
continuous classical-like nonlinear phase-locking (autoresonance). Two-level arguments are used to study the ladder-climbing process, and semiclassical theory describes the autoresonance effect. The regimes of
efficient excitation in the problem are identified and characterized in
terms of three dimensionless parameters describing the driving strength, the
dispersion nonlinearity, and the Kerr-type nonlinearity, respectively. The nonlinearity alters the borderlines between the regimes, and their characteristics.
\end{abstract}

\maketitle


\affiliation{Racah Institute of Physics, Hebrew University of Jerusalem,
Jerusalem 91904, Israel}

\section{Introduction}

\label{introduction}

The discrete nonlinear Schrodinger equation (DNLSE) is an important
nonlinear lattice model describing the dynamics of many systems. Although it
was originally proposed for a biological system \cite{Davydov}, nowadays the
most important of those systems are in the fields of atomic physics and
optics (for a comprehensive review see Ref. \cite{DNLSbook}). Well known examples
analyzed using the DNLSE include bright and dark solitons \cite{WA1,WA2},
Bloch oscillations \cite{WA3} and Anderson localization \cite{WA4} in
optical waveguide arrays. Furthermore, Bloch oscillations \cite{BEC1},
dynamical transitions \cite{BEC2,BEC3}, quantum phase transitions \cite{BEC4,BEC4a}, controlled tunneling \cite{BEC4b,BEC4c,BEC4d}
, and discrete breathers \cite{BEC5,BEC6} were studied in Bose-Einstein condensates (BEC)
in optical lattices.

Due to its prevalence across many fields of research, the ability to
control, excite, and manipulate systems described by the DNLSE is of great
interest. This paper will explore the effects of a chirped frequency
parametric driving added to the DNLSE. Various physical systems including
atoms and molecules \cite{Ex1,Ex2,Ex3,Ex4,Ex5}, anharmonic oscillators \cite%
{AR}, Josephson junctions \cite{Ex6}, plasma waves \cite{Ex7,Ex8}, cold
neutrons \cite{Ex9}, and BEC's \cite{Batalov} all exhibit distinct classical
and quantum mechanical responses to such chirped driving. The classical
response, known as autoresonance (AR) \cite{AR} is characterized by
sustained phase-locking between the system and the drive, yielding
continuing excitation in many dynamical and wave systems. The quantum
mechanical response in the same chirped-drive systems, on the other hand,
is characterized by successive Landau-Zener (LZ) transitions \cite{LZ1,LZ2}
yielding climbing up the energy ladder and hence dubbed quantum ladder
climbing (LC).

But are the AR and LC processes, previously identified in dynamical problems
and continuous wave equations, relevant to the chirped drive discrete
equation in hand? Although different types of chirped drives were studied in
the past in the context of the DNLSE \cite{ChirpLazarSegev,Chirp1,Chirp2},
those works did not study both the quantum mechanical and classical
responses of the same system (in some cases because the system contained too
few sites to study classical-like behavior). This paper will show that
both the quantum mechanical LC and the classical AR could appear in the
chirped drive DNLSE under different choices of parameters. It will explore
the characteristics of both AR and LC processes in the case of DNLSE
with focusing nonlinearity, find the regions in the parameters space where
these processes exist, and demonstrate the degree of control they can exert.

The scope of the paper is as follows: Sec. \ref{Sec1}, introduces the
model and its parametrization. Section \ref{Sec2} is dedicated to the
studying of the periodic DNLSE with periodicity length $N$ of $2$ sites,
demonstrating the quantum-mechanical LZ transitions and the effect of the
explicit Kerr-type nonlinearity. Using this two-level description as a
building block, Sec. \ref{Sec3} characterizes the AR and LC responses when $%
N $ is large, including separation between the regimes in the associated
parameters space. Our conclusions are summarized in Sec. \ref{summary}.

\section{The Model And Parametrization}

\label{Sec1} This paper focuses on a periodic, chirped-drive DNLSE of
the form:
\begin{equation}
i\frac{d\psi _{n}}{dt}+\frac{\left( \psi _{n+1}+\psi _{n-1}-2\psi
_{n}\right) }{\Delta ^{2}}+\left[ \beta \left\vert \psi _{n}\right\vert
^{2}+\varepsilon \cos \phi _{n}\right] \psi _{n}=0,  \label{DNLSE}
\end{equation}%
where $\psi _{n+N}=\psi _{n}$, $\phi _{n}=\frac{2\pi n}{N}-\theta _{d}(t)$, $%
\theta _{d}$ is the driving phase having slowly varying (chirped) frequency $%
\omega _{d}(t)=d\theta _{d}/dt$, we assume $\beta >0$ (focusing Kerr-type
nonlinearity) and initial driving time $t=0$. In the context of the BEC in optical lattices such parametric driving could be realized by spatial and temporal modulation of the lattice, similar to Ref. \cite{BEC4a}. Our proposed driving was
studied in the past without the chirp \cite{SimilarDrive} and is designed to
drive the system between the modes set by the traveling-wave solutions of
the linearized, unperturbed ($\beta ,\varepsilon =0$) equation:
\begin{eqnarray}
\Psi _{n}^{m} &=&\frac{1}{\sqrt{N}}\exp \left( ik_{m}n-iw_{m}t\right) ,
\notag \\
k_{m} &=&\frac{2\pi m}{N},  \label{modes} \\
w_{m} &=&\frac{4}{\Delta ^{2}}\sin ^{2}\left( k_{m}/2\right) ,  \notag \\
m &=&0,1,...,N-1.  \notag
\end{eqnarray}%
It will also be demonstrated below that our results are not limited to this
specific choice of chirped frequency driving and that other driving schemes
could be analyzed in a similar fashion. A particular example is presented in
Appendix \ref{AppA} for zero boundary conditions ($\psi _{0}=\psi _{N-1}=0$).

To proceed, one assumes a constant driving frequency chirp rate $\alpha $,
(i.e., $\theta _{d}=\alpha t^{2}/2$) and uses normalization $%
\sum_{n}\left\vert \psi _{n}\right\vert ^{2}=1$. One can identify four time
scales in the problem: the frequency sweeping time scale $t_{s}=1/\sqrt{%
\alpha }$, the driving time scale $t_{d}=2/\varepsilon $, the characteristic
frequency dispersion time scale $t_{c}=\Delta ^{2}N^{2}/4\pi ^{2}\approx
1/\omega _{1}$ and the Kerr-type nonlinearity time scale $t_{nl}=N/\beta $.
The choice of $t_{nl}$ reflects the effective average value of the Kerr-type
interaction, which is smaller by a factor of $1/N$ than $\beta $ due to our
normalization. Using these four time scales one can define three
dimensionless parameters
\begin{eqnarray*}
P_{1} &=&\frac{t_{s}}{t_{d}}=\frac{\varepsilon }{2\sqrt{\alpha }}, \\
P_{2} &=&\frac{t_{s}}{t_{c}}=\frac{4\pi ^{2}}{\Delta ^{2}N^{2}\sqrt{\alpha }}%
, \\
P_{3} &=&\frac{t_{s}}{t_{nl}}=\frac{\beta }{N\sqrt{\alpha }}.
\end{eqnarray*}%
These parameters characterize the driving strength, the dispersion
nonlinearity, and the Kerr-type nonlinearity, respectively, and fully
determine the evolution of the driven system, as can be seen if one rewrites
Eq. (\ref{DNLSE}) in the dimensionless form:%
\begin{equation}
\begin{array}{ll}
i\frac{d\psi _{n}}{d\tau } & +\frac{N^{2}}{4\pi ^{2}}P_{2}\left( \psi
_{n+1}+\psi _{n-1}-2\psi _{n}\right) \\
& +\left( NP_{3}\left\vert \psi _{n}\right\vert ^{2}+2P_{1}\cos \phi _{n}%
\right) \psi _{n}=0,%
\end{array}
\label{DNLSE1}
\end{equation}%
where $\tau =\sqrt{\alpha }t$ is the dimensionless slow time.

It is convenient at this stage to expand $\psi _{n}=\sum_{m}a_{m}\Psi
_{n}^{m}$ in terms of the linear modes and rewrite (\ref{DNLSE1}) as

\begin{equation}
i\sum_{m}\frac{da_{m}}{d\tau }\Psi _{n}^{m}+NP_{3}K+P_{1}\sum_{m}\left(
e^{i\phi _{n}}+e^{-i\phi _{n}}\right) a_{m}\Psi _{n}^{m}=0,  \label{ModeEq1}
\end{equation}%
where
\begin{equation*}
K=\sum_{m,m^{\prime },m^{\prime \prime}}a_{m^{\prime }}a_{m^{\prime \prime}}^{\ast }a_{m}\Psi
_{n}^{m^{\prime }}\Psi _{n}^{m^{\prime \prime}\ast }\Psi _{n}^{m}.
\end{equation*}%
Next, one combines all $n$ dependent components in the driving term
and in $K$ into a single base function, multiplies Eq. (\ref%
{ModeEq1}) by $\Psi _{n}^{l\ast }$, and sums the result over $n$ using the
orthonormality $\sum_{n}\Psi _{n}^{m}\Psi _{n}^{m^{\prime }\ast }=\delta
_{m,m^{\prime }}$ to get:
\begin{equation}
i\frac{da_{l}}{d\tau }+P_{3}K_{l}+P_{1}[a_{l-1}e^{i\left( \Delta \omega
_{l}\tau -\theta _{d}\right) }+a_{l+1}e^{-i\left( \Delta \omega _{l+1}\tau
-\theta _{d}\right) }]=0.  \label{ModeEq2}
\end{equation}%
Here $\Delta \omega _{l}=\omega _{l}-\omega _{l-1}$, $\omega _{l}=w_{l}/%
\sqrt{\alpha }$ is the dimensionless form of $w_{l}$ and
\begin{equation*}
K_{l}=\sum_{m^{\prime },m^{\prime \prime}}a_{m^{\prime }}a_{m^{\prime \prime}}^{\ast }a_{l-m^{\prime
}+m^{\prime \prime}}e^{i\left( \omega _{m^{\prime }}-\omega _{m^{\prime \prime}}+\omega _{l-m^{\prime
}+m^{\prime \prime}}-\omega _{l}\right) \tau }.
\end{equation*}

Equation (\ref{ModeEq2}) is still exact, and some approximations are needed
to advance the analysis. This is performed by moving to the frame of reference
rotating with the drive and neglecting rapidly oscillating components in $%
K_{l}$. For the stationarity of the terms in $K_{l}$ in the rotating frame
of reference, the phases in the exponents must vanish. Aside from esoteric
examples \cite{SideNote1} this could only be achieved when either $m^{\prime
}=m^{\prime \prime }$ or $m^{\prime }=l$, which after the summation results
for both cases in $\sum_{m}\left\vert a_{m}\right\vert ^{2}a_{l}=a_{l}$, but
the term $\left\vert a_{l}\right\vert ^{2}a_{l}$ is counted twice.
Therefore, in the rotating wave approximation (RWA), one has:
\begin{equation*}
K_{l}\approx 2a_{l}-\left\vert a_{l}\right\vert ^{2}a_{l}.
\end{equation*}%
Finally, one defines $b_{l}=a_{l}\exp \left( il\theta _{d}-i\omega _{l}\tau
-i2\tau \right) $ to get:
\begin{equation}
i\frac{db_{l}}{d\tau }=-b_{l}\left( l\omega _{d}-\omega _{l}\right)
+P_{3}\left\vert b_{l}\right\vert ^{2}b_{l}-P_{1}\left(
b_{l-1}+b_{l+1}\right) ,  \label{ModeEqFinal}
\end{equation}%
where the dimensionless form of $\omega _{d}$ equals $\tau $. It should be
noted that the symmetry $a_{-1}=a_{N-1}$ is broken in system (\ref%
{ModeEqFinal}), as $b_{-1}\neq b_{N-1}$, and therefore a phase factor must
be added to the couplings between modes $0$ and $N-1$. For the sake of this
paper it is sufficient to neglect these couplings, as they are nonresonant at
times $\tau >0$ studied below.

\begin{figure}[]
\includegraphics[width=3.375in]{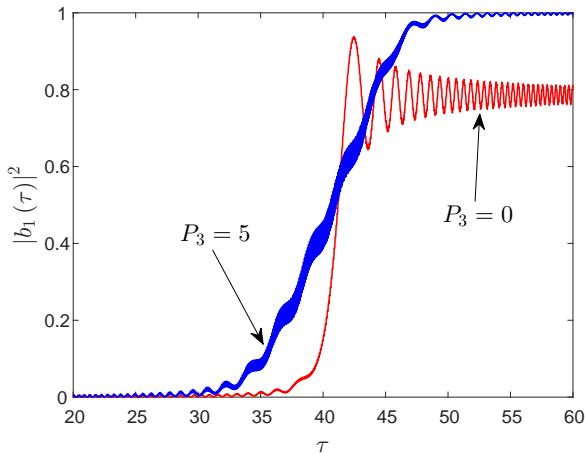}
\caption{The numerical solution of Eq. (\protect\ref{DNLSE1}) for the
population of mode $1$ versus time. The parameters are $N=2$, $P_{1}=0.5$, $%
P_{2}=100$ and $P_{3}=0$ (red) or $5$ (blue).}
\label{Fig1}
\end{figure}

Equation (\ref{ModeEqFinal}) can yield complex dynamics depending on the
parameters of the problem. Even the very basic example of $N=2$ illustrated
in Fig. \ref{Fig1} exhibits remarkably different evolutions when only
parameter $P_{3}$ is changed. Therefore, Sec. \ref{Sec2} will discuss the $N=2$ case first. Naturally, such a system can not
exhibit classical-like behavior involving many modes, but it provides key
insights into the two-level interactions which will be used in Sec. \ref%
{Sec3} in studying the $N\gg 1$ case.

\section{$N=2$ Case}

\label{Sec2}

We write Eq. (\ref{ModeEqFinal}) explicitly for $N=2$:
\begin{equation}
i\frac{d}{d\tau }\left(
\begin{array}{c}
b_{0} \\
b_{1}%
\end{array}%
\right) =\left(
\begin{array}{cc}
P_{3}\left\vert b_{0}\right\vert ^{2} & -P_{1} \\
-P_{1} & P_{3}\left\vert b_{1}\right\vert ^{2}-\tau +\omega _{1}%
\end{array}%
\right) \left(
\begin{array}{c}
b_{0} \\
b_{1}%
\end{array}%
\right) .  \label{2Level}
\end{equation}%
As mentioned in Sec. \ref{Sec1} the couplings $b_{0}\leftrightarrow b_{-1}$
and $b_{1}\leftrightarrow b_{2}$ in Eq. (\ref{ModeEqFinal}) are nonresonant,
and thus neglected in Eq. (\ref{2Level}).

In the linear case, $P_{3}=0$, Eq. (\ref{2Level}) takes the well-known LZ
form \cite{LZ1,LZ2} with an avoided energy crossing at $\tau _{c}=\omega
_{1} $ \cite{SideNote2}. If one starts in the ground state, $\left\vert
b_{0}\left( \tau =0\right) \right\vert =1,$ the fraction of the population
transferred to mode $1$ is given by the LZ formula $\left\vert b_{1}\left(
\tau \gg \tau _{c}\right) \right\vert ^{2}=1-\exp \left( -2\pi
P_{1}^{2}\right) $ \cite{LZ1,LZ2}. The red curve in Fig. \ref{Fig1} shows an
example for such LZ dynamics for $P_{1}=0.5$ and $P_{2}=100$. One can see a
rapid population transfer around $\tau _{c}\approx 40.5$ converging to the
value given by the LZ formula. However, when the explicit Kerr-type
nonlinearity is introduced, the dynamics changes significantly. This is
shown by the blue curve of Fig. \ref{Fig1}, where $P_{3}=5$, whereas all other
parameters are the same. In this case, the population transfer is much
slower and almost linear in time reaching a higher final state for the same
driving parameter $P_{1}$.

\begin{figure}[]
\includegraphics[width=3.375in]{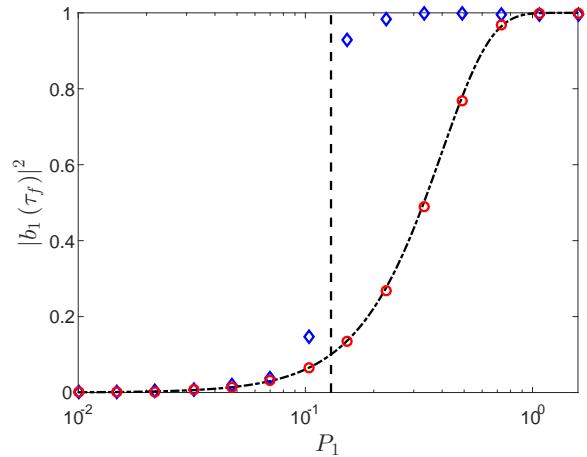}
\caption{The numerical solution of Eq. (\protect\ref{DNLSE1}) for the final
population of mode $1$ as function of $P_{1}$. The parameters are $N=2$, $%
P_{2}=\protect\tau_{f}=100$ and $P_{3}=0$ (red circles) or $5$ (blue
diamonds). The dashed vertical line shows the theoretical NLZ threshold [Eq.
\protect\ref{NLZThreshold}], whereas the dashed-dotted curve is the theoretical LZ
formula.}
\label{Fig2}
\end{figure}

Figure \ref{Fig2} shows the final population of mode $1$ at $\tau _{f}=100$
as a function of $P_{1}$ and further demonstrates the differences between
the two scenarios. In the linear $P_{3}=0$ case (red circles), the population
transfer follows the LZ formula (dashed-dotted curve), whereas for $P_{3}=5$ (blue
diamonds) the population of mode $2$ "jumps" abruptly, reaching nearly full
population transfer at lower driving strengths than in the linear case. This
so-called nonlinear Landau-Zener transition (NLZ) was studied in the past in
various contexts \cite{Batalov,ChirpLazarSegev,NLZ1,NLZ2,NLZ3}. It was shown that
the growth of the population of mode $2$ is in fact linear in time (with
superimposed oscillations), as illustrated in Fig. \ref{Fig1}, and a nearly
full population transfer takes place if $P_{1}$ exceeds a sharp threshold
\cite{NLZ1,ChirpLazarSegev,Batalov}:
\begin{equation}
P_{1,cr}^{NLZ}\approx 0.29/\sqrt{P_{3}}.  \label{NLZThreshold}
\end{equation}%
The value of $P_{1,cr}^{NLZ}$ is shown in Fig. \ref{Fig2} by vertical dashed
line, in good agreement with the numerically observed "jump" in the
transfer of population.

\begin{figure}[]
\includegraphics[width=3.375in]{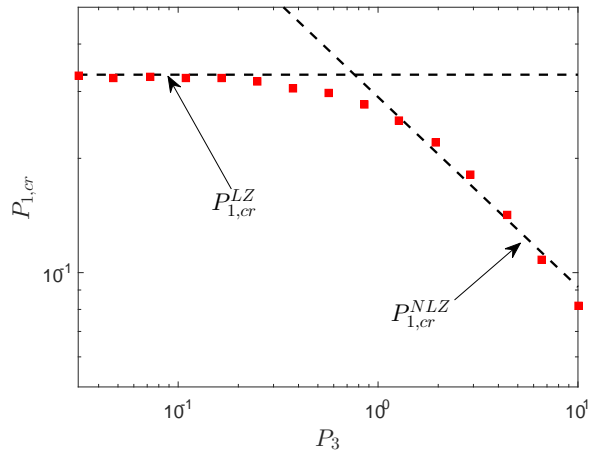}
\caption{The solution of Eq. (\protect\ref{DNLSE1}) for the threshold value
$P_{1,cr}$ yielding transfer of one half of the population to mode $1$,
as a function of $P_{3}$. The parameters are $N=2$, $P_{2}=\protect\tau%
_{f}=200 $, and the dashed lines show the theoretical predictions according
to the LZ formula and Eq. (\protect\ref{NLZThreshold}). Numerical
uncertainty is smaller than the marker sizes.}
\label{Fig3}
\end{figure}

One can further demonstrate the differences between LZ and NLZ regimes by
defining $P_{1,cr}$ as the value of $P_{1}$ for which half of the population
transitions from mode $0$ to mode $1$. The numerically obtained value of $%
P_{1,cr}$ is plotted in Fig. \ref{Fig3} versus $P_{3}$. For large enough $%
P_{3}$, $P_{1,cr}$ matches $P_{1,cr}^{NLZ}$ (dashed diagonal line). However,
in the LZ regime the LZ formula yields $P_{1,cr}^{LZ}=\sqrt{-\ln 0.5/2\pi }%
\approx 0.33$. And, indeed, for low $P_{3}$, $P_{1,cr}$ matches $%
P_{1,cr}^{LZ}$ (dashed horizontal line). The intersection of the two
threshold values $P_{1,cr}^{NLZ}=P_{1,cr}^{LZ}$ yields a good estimate for
the value of $P_{3}$ for which the transition between the two regimes takes
place.

Our driving perturbation differs from that assumed in the asymptotic
theories of LZ and NLZ processes because it involves a finite driving time
prior to the energy crossing at $\tau _{c}$. Nevertheless, it will be assumed
that $\tau _{c}$ is large enough for the two theories to be valid, which can
always be accomplished by increasing $P_{2}$ (as $\tau _{c}\propto P_{2}$).
Nevertheless, the breaking of this assumption is important in studying the $N\gg
1$ case in Sec. \ref{Sec3} and, thus, requires a further discussion. For $%
\tau _{c}$ to be large enough for the applicability of the asymptotic LZ and
NLZ theories, it must be larger than the characteristic time of population
transfer from one mode to the next. In the case of LZ, the transition time $%
\Delta \tau _{LZ}$ is of order $O(1)$ when $P_{1}$ is small and $O(P_{1})$
when it is large, therefore we estimate $\Delta \tau _{LZ}=1+P_{1}$ \cite%
{LZtimescales}. In the case of NLZ the estimate is $\Delta \tau
_{NLZ}=2P_{3} $ \cite{Batalov}. These two times can be combined into a
single estimate for the transition duration
\begin{equation}
\Delta \tau =1+P_{1}+2P_{3}  \label{deltatau}
\end{equation}%
and, therefore, $\tau _{c}\gg \Delta \tau $ guarantees that the dynamics is
of the asymptotic LZ or NLZ type. Furthermore, since the neglected terms in
the derivation of Eq. (\ref{ModeEqFinal}) and Eq. (\ref{2Level}) oscillate
with frequency proportional to $P_{2}$, the aforementioned condition also
justifies the RWA approximation.

\section{Quantum and Classical Effects for Large $N$}

\label{Sec3}

\begin{figure*}[t]
\includegraphics[width=6.75in]{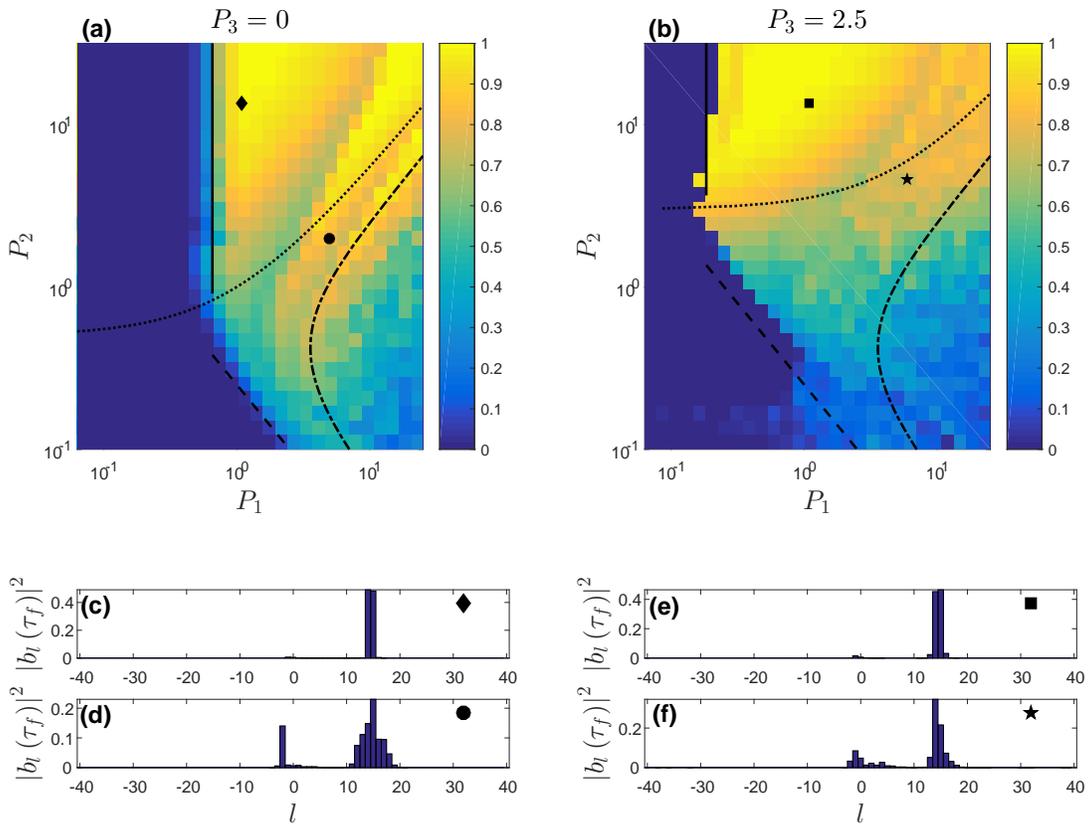}
\caption{Color coded excitation efficiencies (see the text) in the $P_{1,2}$
parameter space for (a) $P_{3}=0$ and (b) $2.5$, as obtained from the
numerical solution of Eq. (\protect\ref{DNLSE1}), with $N=80$ and $\protect%
\tau_{f}=\protect\tau_{15}\approx 23.1P_{2}$. The lines represent different borderlines in the parameter space - The efficient LC threshold (solid line), efficient AR threshold (dashed line), quantum-classical separation (dotted line), and the large separatrix boundary (dashed-dotted line).
 Panels (c)-(f) show the
population of each mode at $\protect\tau_{f}$ for $P_{1,2}$ values shown by
the corresponding markers in panels (a) and (b). For clarity, modes with $l>N/2$
are shifted and presented as $l<0$.}
\label{Fig4}
\end{figure*}

The controlled excitation in our system is not limited to the $N=2$ case, therefore
the $N\gg 1$ limit is considered next (for some remarks on the case of moderate $N$
see App. \ref{AppB}). Panels (c)-(e) in Fig. \ref{Fig4} show histograms of
the final populations $\left\vert b_{l}\left( \tau _{f}\right) \right\vert
^{2}$ for $N=80$ and $\tau _{f}\approx 23.1P_{2}$. The parameters $P_{1,2}$
in these panels correspond to those shown by corresponding markers in the parameter space
of panels (a) and (b) , where $P_{3}=0$ and $2.5$, respectively. These
figures illustrate a controlled transfer of the populations to the vicinity
of a target mode (in this case, $l\approx 15$), with some width around this
mode. In this section we show how the different parameters in the
problem control the target mode, the fraction of the excited population, and
the width of the excited distribution of modes.

\subsection{Quantum-mechanical ladder climbing}

Panels (c) and (e) in Fig. \ref{Fig4} exhibit very narrow distributions (1 to 2
modes) and hint at the connection between the cases of $N=2$ and $N\gg 1$.
This connection becomes apparent when one examines only two mode interaction $%
l-1\leftrightarrow l$ and neglects other modes in Eq. (\ref{ModeEqFinal}),
i.e. solves

\begin{equation}
i\frac{d}{d\tau }\left(
\begin{array}{c}
b_{l-1} \\
b_{l}%
\end{array}%
\right) =\left(
\begin{array}{cc}
\Gamma _{l-1} & -P_{1} \\
-P_{1} & \Gamma _{l}%
\end{array}%
\right) \left(
\begin{array}{c}
b_{l-1} \\
b_{l}%
\end{array}%
\right) ,  \label{2LevelN}
\end{equation}%
where $\Gamma _{l}=P_{3}\left\vert b_{l}\right\vert ^{2}-l\tau +\omega _{l}$%
. Similar to the case of $N=2$, Eq. (\ref{2LevelN}) takes the form of LZ or
NLZ transition, depending on the value $P_{3}$. However, in this case, there
are many such transitions (resonances) and their timing is $l$ dependent.
This temporal separation between the transitions allows the system to
successively perform quantum energy LC via pairwise LZ or
NLZ transitions. The time $\tau _{l}$ of the transition $l-1\leftrightarrow
l $ can be found by equating $\Gamma _{l-1}=\Gamma _{l}$ (energy crossing)
which yields

\begin{equation}
\tau_{l}=\frac{P_{2}N^2}{\pi^{2}}\sin\left({\frac{\pi}{N}}\right)\sin\left(%
\frac{\pi\left[2l-1\right]}{N}\right).  \label{taul}
\end{equation}

Examining Eq. (\ref{taul}), one can identify a resonant pathway of
consecutive transitions from the ground state to $l\approx N/4$. The final
driving time $\tau _{f}$ dictates how high in $l$ the system will climb and
sets the target mode for the process. In the simulations of Fig. \ref{Fig4},
$\tau _{f}\approx 23.1P_{2}$ so that $\tau _{f}=\tau _{15}$, as
could be observed in panels (c)-(f). If the consecutive transitions are well
separated in time, one can treat them as individual LZ or NLZ transitions,
and use all of the results discussed in Sec. \ref{Sec2} for $N=2$.
Specifically, the probability of population transfer will follow the LZ
formula and will exhibit a sharp threshold on $P_{1}$ for the NLZ
transition. Thus, the excitation efficiency (the fraction of the excited
population) in the two cases should exhibit different characteristics. Once
again, one can define $\overline{P}_{1,cr}$ as the value of $P_{1}$, which
will drive $50\%$ of the population after $r$ transitions. Using the LZ
formula one can calculate

\begin{equation}
\overline{P}_{1,cr}^{LZ}=\sqrt{-\frac{\ln \left( 1-2^{-1/r}\right) }{2\pi }}.
\label{LZThresholdN}
\end{equation}

For NLZ transitions, the sharp threshold guarantees that if the first
transition was efficient, it will continue to be efficient later and, thus,

\begin{equation}
\overline{P}_{1,cr}^{NLZ}=P_{1,cr}^{NLZ}.  \label{NLZThresholdN}
\end{equation}%
To check this prediction, Eq. (\ref{DNLSE1}) was solved numerically with $%
N=80$. The excitation efficiency was defined as the total population between
modes $10$ and $20$ (upper half of the resonantly accessible modes). These
results are color coded in panels (a) and (b) of Fig. \ref{Fig4}. The
population undergoes $r=10$ transitions between the ground state and the
measurement window, and the corresponding $P_{1,cr}$ according to Eqs. (\ref%
{LZThresholdN}), and (\ref{NLZThresholdN}) is plotted as vertical solid
lines in panels (a) and (b). One can see that for large enough $P_{2}$, the
excitation efficiency grows as expected with $P_{1}$: it significantly
increases in the vicinity of $P_{1,cr}$, and grows sharply in the NLZ case
[panel (b)].

The agreement with the numerics for high enough $P_{2}$ only is expected, as
the assumption that different transitions are well separated in time, is not valid for small $P_{2}$. Using the logic of Sec. \ref{Sec2}, for
the transitions to be well separated, one must require the typical time between
the transitions to be larger than the typical duration of a single transition,
as given by Eq. (\ref{deltatau}). In the limit $N\gg 1$, $l\ll N$, Eq. (\ref%
{taul}) shows that the time between two successive transitions is $2P_{2}$ (regardless of the value of $P_{3}$, since the temporal separation is set by the linear unperturbed problem $P_{1}=P_{3}=0$)
and, therefore,

\begin{equation}
P_{2}\gg \frac{1}{2}+\frac{P_{1}}{2}+P_{3},  \label{separation}
\end{equation}%
is the criterion for the LC. The line in the $P_{1,2}$ space on which the
two sides of inequality (\ref{separation}) are equal is shown by the dotted lines in panels (a) and (b) of Fig. \ref{Fig4}. One can see that the LC
prediction holds only above this line. Furthermore, panels (c) and (e) of Fig. \ref{Fig4} (corresponding to final simulation time and parameters in the LC regime) involve only two levels as expected from separated successive LZ transitions. A movie illustrating this dynamics at earlier times for the parameters of panel (c) is presented in the Supplemental Material \cite{SUP}. The observed temporal separability of the transitions differs from the lack of separability in the context of counterdiabatic protocols \cite{CA1}.

It should be noted that, although initially the transitions are nearly evenly separated (similar to other LC
systems \cite{Ex3,Ex5,LZtimescales}) as one approaches larger $l$, the
transitions become more frequent. Condition (\ref{separation}) does not hold
in this case, and the dynamics will cease to be of LC nature. However, as
could be observed in Fig. \ref{Fig4} and will be discussed below, condition (%
\ref{separation}) is still sufficient in the context of excitation
efficiency.

But what happens when criterion (\ref{separation}) is not met and the transitions are not well separated? Figure \ref%
{Fig4} shows that there could still be efficient excitation, but now many
modes are coupled at a time. This mixing of many different modes leads to
classical-like behavior. This is also hinted by the wide distributions
observed in panels (d) and (f), where the parameters are outside the LC regime.
The semiclassical analysis of this regime will be our next goal.

\subsection{Semiclassical autoresonant regime}

For studying the semiclassical evolution of the system when condition (\ref%
{separation}) is not met, return to Eq. (\ref{DNLSE1}) and assume that
this set can be replaced by a continuous equation in the limit $N\gg 1$.
Then one expands
\begin{equation*}
\psi ^{n\pm 1}=\sum_{j=0}^{\infty }\frac{1}{j!}\frac{d^{j}\psi ^{n}}{dn^{j}}%
\left( \pm 1\right) ^{j},
\end{equation*}%
inserts this expansion into Eq. (\ref{DNLSE1}) and defines the continuous
space-like variable $x\equiv n$ to get
\begin{equation}
i\frac{\partial \psi }{\partial \tau }+P_{2}\frac{N^{2}}{2\pi ^{2}}%
\sum_{j=1}^{\infty }\frac{1}{\left( 2j\right) !}\frac{\partial ^{2j}\psi }{%
\partial x^{2j}}+(NP_{3}\left\vert \psi \right\vert ^{2}+2P_{1}\cos \Phi
)\psi =0.  \label{WKB1}
\end{equation}%
Here, $\psi =\psi \left( x,\tau \right) $ and $\Phi =k_{0}x-\theta _{d}$
with $k_{0}=2\pi /N$. At this point, one writes the wave-like eikonal ansatz $%
\psi =b\left( x,\tau \right) \exp \left[ iS\left( x,\tau \right) \right] $
\cite{Tracy}, where $S$ is viewed as a rapidly oscillating phase variable,
whereas $b$ is a slow amplitude. In addition, it is assumed that the derivatives of the
fast phase
\begin{eqnarray*}
k &\equiv &\frac{\partial S}{\partial x}, \\
\Omega  &\equiv &-\frac{\partial S}{\partial \tau }
\end{eqnarray*}%
are both slow. The slowness in our problem means $%
|\partial (\ln G)/\partial x|\ll k$, where $G$ is any of the slow variables
above \cite{Tracy}. The eikonal ansatz models our basis modes $\Psi _{n}^{m}$
in discrete formalism. For example, the increase in $k$ in time would
describe a transition to higher modes. Next, one approximates $\frac{d^{2j}\psi }{dx^{2j}}\approx
be^{iS}\left( ik\right) ^{2j}$ (neglecting small derivatives of $b$ and $k$%
), inserts this approximation into Eq. (\ref{WKB1}) and identifies the sum over
$j$ as the Taylor expansion of $-2\sin ^{2}\left( k/2\right) $ to obtain
\begin{equation}
i\frac{db}{d\tau }+b\Omega -P_{2}\frac{N^{2}}{\pi ^{2}}b\sin ^{2}\frac{k}{2}%
+(NP_{3}b^{2}+2P_{1}\cos {\Phi )}b=0.  \label{WKB2}
\end{equation}%
The imaginary part of Eq. (\ref{WKB2}) yields $\frac{db}{d\tau }=0$. For a
more accurate description of the evolution of the amplitude $b$ in the
eikonal ansatz, one must go to a higher order of the approximation. However,
it can be shown that the essentials of the resonant dynamics can be revealed
without resolving $b$. We start with the case $P_{3}=0$ for which the real
part of Eq. (\ref{WKB2}) reads
\begin{equation}
\Omega \left( x,\tau \right) =P_{2}\frac{N^{2}}{\pi ^{2}}\sin ^{2}\frac{%
k\left( x,\tau \right) }{2}-2P_{1}\cos {\Phi }.  \label{Omega}
\end{equation}

Equation (\ref{Omega}) is a first order partial differential equation for
the phase variable $S$ in the eikonal ansatz, and can be solved along
characteristics (rays). To this end, Eq. (\ref{Omega}) can be interpreted as
defining the function of three variables $\Omega =\Omega \left( x,k,\tau
\right) ,$ where $k$ is also a function of $x,t$ and introduce the
characteristics via
\begin{equation}
\frac{dx}{d\tau }=\frac{\partial \Omega \left( x,k,\tau \right) }{\partial k}%
.  \label{1}
\end{equation}%
Note that by construction,
\begin{equation*}
\frac{d\Omega }{dx}+\frac{\partial k}{\partial \tau }=0,
\end{equation*}%
which can be rewritten as

\begin{equation*}
\frac{\partial \Omega }{\partial x}+\frac{\partial \Omega }{\partial k}\frac{%
\partial k}{\partial x}+\frac{\partial k}{\partial \tau }=0.
\end{equation*}%
This yields the second ray equation%
\begin{equation}
\frac{dk}{d\tau }=\frac{\partial k}{\partial \tau }+\frac{dx}{d\tau }\frac{%
\partial k}{\partial x}=-\frac{\partial \Omega }{\partial x},  \label{2}
\end{equation}%
which, in combination with (\ref{1}), provides a complete system for
following $x$ and $k~$along the rays. Note that these two equations comprise
a Hamiltonian set with $\Omega \left( x,k,\tau \right) $ being the
Hamiltonian. In addition,
\begin{equation}
\frac{d\Omega }{d\tau }=\frac{\partial \Omega }{\partial \tau }  \label{3}
\end{equation}%
and
\begin{equation}
\frac{dS}{d\tau }=\frac{\partial S}{\partial \tau }+\frac{\partial S}{%
\partial x}\frac{dx}{d\tau }=-\Omega +k\frac{\partial \Omega }{\partial k}.
\label{4}
\end{equation}%
Equations (\ref{1})-(\ref{4}) can be conveniently solved to provide the
phase factor $S$ as well as $x,k$, and $\Omega $ along the rays, provided
the initial condition $S(x,\tau =0)$ is known on some interval of $x$. This
knowledge also yields the initial conditions $k(x,\tau =0)$ and $\Omega
(x,\tau =0)$ [from (\ref{Omega})] on this interval and solving the system (\ref{1})-(\ref{4}) by
starting on the interval allows to evolve the system in time. However,
analyzing the phase-space of our Hamiltonian set is just as informative as shown below.

We insert Eq. (\ref{Omega}) into Eqs. (\ref{1}) and (\ref{2}) and recall that $%
\Phi =k_{0}x-\tau ^{2}/2$ to get

\begin{eqnarray}
\frac{d\Phi }{d\tau } &=&P_{2}\frac{N}{\pi }\sin k-\tau ,  \label{HamEqsA} \\
\frac{dk}{d\tau } &=&-P_{1}\frac{4\pi }{N}\sin \Phi .  \label{HamEqsB}
\end{eqnarray}%
This system has the form known from many other classical autoresonantly
driven systems studied in the past (e.g. Refs. \cite{Ex5,PhaseSpace}), so
previously known results can be used directly in our case and we briefly
describe these results. The angle $\Phi $ acts as a phase-mismatch between
the driving force and the system. When the resonance condition $\frac{d\Phi
}{d\tau }\approx 0$ is met continuously, $P_{2}\frac{N}{\pi }\sin k$ follows
the driving frequency ($\omega _{d}=\tau $), thus the system is driven to
higher modes. It should be noted that this resonance condition is identical
to that given by Eq. (\ref{taul}) in the limit $N,l\gg 1$. Next, we take
the second derivative of (\ref{HamEqsA}) and insert (\ref{HamEqsB}) to get

\begin{equation}
\frac{d^{2}\Phi }{d\tau ^{2}}=-4P_{1}P_{2}\cos k\sin \Phi -1.
\label{TiltPend}
\end{equation}%
Here, we approximate $k\approx k_{r}$, where $k_{r}\left( \tau \right) $ is
the value of $k$ satisfying the exact resonance condition \cite%
{Ex5,PhaseSpace}. Then, Eq. (\ref{TiltPend}) describes a pendulum with a
time varying frequency and under the action of a constant torque. If $%
4P_{1}P_{2}\cos k_{r}>1$, the phase-space of the system has both open and
closed trajectories. On the open trajectories, $\Phi $ grows indefinitely
and $\sin k$ does not follow the driving frequency. In contrast on the
closed trajectories, $\Phi $ and $d\Phi /d\tau $ are bounded and yield
sustained phase-locking (autoresonance) of the system to the drive, i.e., a
continuing excitation of $k$. The separatrix is the trajectory separating
the closed and open trajectories in phase-space, and it only exists if $%
4P_{1}P_{2}\cos k_{r}>1$. Therefore, if one takes $\cos k_{r}$ at its
maximal value of $1$, one obtains the threshold

\begin{equation}
P_{1}P_{2}=\frac{1}{4},  \label{ClTh}
\end{equation}%
below which no autoresonant excitation is possible. This threshold is shown
by the diagonal dashed lines in panels (a) and (b) in Fig. \ref{Fig4}, showing
good agreement with the numerical simulations for both values of $P_{3}$
\cite{SideNote3}, even though we have assumed $P_{3}=0$ above. This can be
explained by observing that when $P_{3}\neq 0$, only Eq. (\ref{HamEqsB}) is
affected and becomes

\begin{equation}
\frac{dk}{d\tau }=-P_{1}\frac{4\pi }{N}\sin \Phi +NP_{3}\frac{\partial
\left( b^{2}\right) }{\partial x}.  \label{dkdtP3}
\end{equation}%
Initially, in our simulations the additional term in Eq. (\ref{dkdtP3})
vanishes since $b$ is independent of $x$. Therefore, initially, the existence of the separatrix is
not affected by $P_{3}$. At later times, if the separatrix exists, the focusing nonlinearity narrows the
distribution and, thus, doesn't scatter the trapped trajectories out of the
separatrix. Numerically, the narrowing of the distribution is seen when
comparing panels (d) and (f) in Fig. \ref{Fig4}. Hence, the initial separatrix governs the existence of trapped trajectories, and since it is independent of $P_{3}$, threshold (\ref{ClTh}) describes the case $P_{3}\neq 0$ as well.

Until now, we have treated the trajectories inside the separatrix as those
which will be excited to large $k$, but this is not the case when the
separatrix becomes too large. In this case, even when a significant portion
of the population is inside the separatrix, not all of it will be excited to
large $k$, and subsequently will be precluded from our numerical
measurement. The dashed-dotted line in panels (a) and (b) of Fig. \ref{Fig4}
marks the values of $P_{1,2}$ for which the separatrix extends in $k$ at
$\tau _{f}$ below our measurement window ($\pi /4$). Below this line the
excitation efficiency drops, as more population ends up outside the
measurement window. The aforementioned narrowing of the autoresonant bunch
hinders this argument for $P_{3}\neq 0$, but nevertheless, for the values of
$P_{3}$ in our simulations, this criterion still qualitatively agrees with
the numerical simulations. The details of the separatrix related
calculations are described in Appendix \ref{AppC}.

Finally, we return to the quantum-classical separation line given by Eq. (%
\ref{separation}), which was derived under the assumption of equidistant
energy crossings. Although this assumption breaks when the population is
transferred to higher modes and several modes are coupled simultaneously,
one can again use the semiclassical arguments as above. The same logic
dictates that the excited population will undergo a dynamical transition
from LC type evolution to AR evolution. This is guaranteed by the population
being in resonance (again, one should note the similarities between the
quantum and the classical resonance conditions), whereas the parameters in the
efficient LC regime are always sufficient for efficient AR.

It should be noted that some features in panels (a) and (b) of Fig. \ref{Fig4} could not be accounted for using the theoretical framework described in this section. For example, the efficiency "dip" close to the quantum-classical separation line (dotted line in the figure) could not be explored using the LC or AR arguments, as both approximations fail in this area of the parameter space. Furthermore, using the semi-classical theory to calculate the expected efficiency in the AR regime of the parameter space is beyond the scope of this paper. The main obstacle is the determination of the proper distribution of initial conditions for Eqs. (\ref{HamEqsA}) and (\ref{HamEqsB}). In a different context this calculation was possible when the system's initial condition was a thermal state rather than the ground state \cite{Ex5}.

\section{Summary}

\label{summary}

In conclusion, we have studied the problem of the resonantly driven discrete
(periodic over $N$ sites) nonlinear Schrodinger equation for a ground state
initial condition. Based on four characteristic time scales in the problem,
we introduced three dimensionless parameters $P_{1-3}$ characterizing the
driving strength, the dispersion nonlinearity, and the Kerr-type
nonlinearity, respectively and analyzed their effects on the resonant
evolution. First, we analyzed the case of $N=2$ and used it to illustrate
and analyze the processes of linear ($P_{3}=0$) and nonlinear ($P_{3}>0$)
Landau-Zener transitions. We have used this two-level description in
generalizing to the case of $N\gg 1$ and showed how successive linear or
nonlinear Landau-Zener transitions, or LC, can occur in
some regions of the three parameters space. Finally, we used semiclassical
arguments to show how in a different region of the parameters space, when the transitions are not well separated and many modes are mixed, the
classical-like AR evolution could appear. Our analysis
identified the key borderlines in the parameter space, including the LC-AR
separation line and the thresholds for effective LC or AR evolution.

The explicit Kerr-type nonlinearity introduces several new effects. First, a single nonlinear Landau-Zener transition is longer than the linear counterpart, and presents a sharp threshold with respect to the driving strength for achieving a full population transfer. As a result, in the case of $N\gg 1$, the LC regime is moved to higher-$P_{2}$ values in the $P_{1,2}$ parameter space. Furthermore, the effective LC threshold becomes sharp and is moved to lower-$P_{1}$ values in the parameter space. However, the efficient AR threshold remains the same and only the width of the autoresonant wavepacket narrows.

The two resonant mechanisms available in the DNLSE allow for intricate
control, manipulation, and excitation of the system, and one can efficiently
excite either a narrow (via LC) or a broad (via AR) distribution around
given target modes. Our analysis was not limited to the case of periodic
boundary conditions. The discussion of similar effects in the DNLSE with
zero boundary conditions was presented in Appendix \ref{AppA}. Furthermore,
we expect that by adjusting the parameters of the problem both temporally
and spatially, one can use the resonant mechanisms studied here to
manipulate the system in the configuration space. In the context of optical
waveguide arrays some of these effects were illustrated previously by
spatially chirping the refractive index of each waveguide \cite%
{ChirpLazarSegev}.

Owing to the versatility of the resonant mechanisms, their appearance for
various initial and boundary conditions, and the relevance of the DNLSE to
many experimental systems (particularly in the field of atomic physics and
optics), this paper may open many new possibilities for future research. It would be also interesting to explore counterdiabatic schemes \cite{CA1,CA2} in this system.

\begin{acknowledgments}
This work was supported by the Israel Science Foundation Grant No. 30/14.
\end{acknowledgments}


\appendix

\section{Zero Boundary Conditions}

\label{AppA} The resonant mechanisms discussed in this paper are not limited
to the setting described in Sec. \ref{Sec1}. As an important additional
demonstration, we will now show how the driven DNLSE with zero boundary
conditions exhibits the same resonant characteristics. To perform this, we return
to Eq. (\ref{DNLSE}), but now imposing $\psi _{0}=\psi _{N-1}=0$ at all
times (reducing the system to $N-2$ degrees of freedom) and using a modified
standing wave-type chirped driving:
\begin{equation}
\begin{array}{ll}
i\frac{d\psi _{n}}{dt} & +\frac{1}{\Delta ^{2}}\left( \psi _{n+1}+\psi
_{n-1}-2\psi _{n}\right) \\
& +\left[ \beta \left\vert \psi _{n}\right\vert ^{2}+\varepsilon \cos \theta
_{d}\cos \left( \frac{\pi n}{N-1}\right) \right] \psi _{n}=0.\label{A1DNLSE}%
\end{array}%
\end{equation}

To replicate the analysis of Sec. \ref{Sec2}, the new basis functions are
the standing wave solutions of the linearized, unperturbed ($\beta
,\varepsilon =0$) equation:
\begin{eqnarray*}
\Psi _{n}^{m} &=&\sqrt{\frac{2}{N-1}}e^{-iw_{m}t}\sin \left( k_{m}n\right) ,
\\
k_{m} &=&\frac{\pi m}{N-1}, \\
w_{m} &=&\frac{4}{\Delta ^{2}}\sin ^{2}\left( k_{m}/2\right) , \\
m &=&1,2,...,N-2.
\end{eqnarray*}%
The fact that the dispersion remains the same for both types of boundary
conditions is important in exhibiting the same resonant characteristics. It
is possible to define the parameters $P_{1-3}$ in much the same way as in
Sec. \ref{Sec1}, but we refrain from this to avoid excessive notations at
this point. We continue, following Sec. \ref{Sec1}, to finding the
corresponding DNLSE for coefficients $a_{m}$ in the expansion $\psi
_{n}=\sum_{m}a_{m}\Psi _{n}^{m}$. Inserting this expansion into Eq. (\ref%
{A1DNLSE}), multiplying by ${\Psi _{n}^{l}}^{\ast }$ and summing over $n$ we
get
\begin{equation}
\begin{array}{ll}
i\frac{da_{l}}{dt} & +\frac{\varepsilon }{2}\cos \theta _{d}\left[
a_{l-1}e^{i\Delta w_{l}t}+a_{l+1}e^{-i\Delta w_{l+1}t}\right] \\
& +\frac{\beta }{2\left( N-1\right) }\left[
-A_{1}^{1}+A_{-1}^{1}+A_{1}^{-1}-A_{-1}^{-1}\right] =0,%
\end{array}%
\end{equation}%
where
\begin{equation*}
A_{j}^{k}=\sum_{m^{\prime },m^{\prime \prime}}a_{l+jm^{\prime }+km^{\prime \prime}}a_{m^{\prime }}^{\ast
}a_{m^{\prime \prime}}e^{-i\left( w_{l+jm^{\prime }+km^{\prime \prime}}-w_{m^{\prime
}}+w_{m^{\prime \prime}}-w_{l}\right) t}.
\end{equation*}%
Now, we employ the RWA to get
\begin{equation*}
-A_{1}^{1}+A_{-1}^{1}+A_{1}^{-1}-A_{-1}^{-1}\approx 3a_{l}-a_{l}\left\vert
a_{l}\right\vert ^{2},
\end{equation*}%
and
\begin{equation*}
\cos \theta _{d}\approx \frac{1}{2}e^{-i\theta _{d}},
\end{equation*}%
for the resonant pathway ascending from mode $0$. Finally, the
transformation to the rotating frame of reference $b_{l}=a_{l}\exp \left(
il\theta _{d}-iw_{l}t-i3t\right) $ yields
\begin{equation}
i\frac{db_{l}}{dt}=-b_{l}\left( l\frac{d\theta _{d}}{dt}-w_{l}\right) +\frac{%
\beta }{2\left( N-1\right) }\left\vert b_{l}\right\vert ^{2}b_{l}-\frac{%
\varepsilon }{4}\left( b_{l-1}+b_{l+1}\right) ,
\end{equation}%
which has the same form as Eq. (\ref{ModeEqFinal}). Therefore, the system
with zero boundary conditions could be controlled and excited in the same
way as the system with periodic boundary conditions. Note that in this case
there is no coupling between modes $1$ and $N-2$, removing some of the
subtleties encountered in the original problem.

\section{Moderate $N$ Case}

\label{AppB} For moderate $N,$ the semiclassical description is not valid,
but one can still induce a ladder-climbing type behavior. However, unlike
the case of $N\gg 1$, now the exact structure of the resonant ladder plays a
more significant role. For example, if $N$ is divisible by $4$ the last two
transitions in the resonant pathway will occur simultaneously resulting in a
three level LZ transition (sometimes referred to as a "bow tie" transition)
\cite{3LZA,3LZB,3LZC}. In this case, the efficiency of this double
transition is given by $\left( 1-\exp \left[ -\pi P_{1}^{2}\right] \right)
^{2}$ \cite{3LZB}. This effect could only (realistically) be observed for
moderate $N$, as for the $N\gg 1$ case, the system will already behave
classically when this final transition is reached.

Although there is no semi-classical dynamics in this case, the separation line
of the form (\ref{separation}) is still useful in demonstrating when the
system could undergo the full ladder-climbing process from mode $0$ to the
maximal accessible mode $l_{max}=D+1$ ($D\ $being $N/4$ rounded down to the
nearest integer). As in Sec. \ref{Sec3}, we must demand that the minimal
time between transitions is longer than the duration of a single transition
as given by Eq. (\ref{deltatau}). One can show that this minimal time is
either the time of the first transition $\tau _{1}$ when $N\leq 4$, or the
time between the two last transitions when $N>4$. The time between the two
last transitions is $\tau _{l_{max}}-\tau _{l_{max}-1}$ (when $N$ is not
divisible by $4$) or $\tau _{l_{max}-1}-\tau _{l_{max}-2}$ (when $N$ is
divisible by $4$).

\section{Separatrix Related Calculations}

\label{AppC} As discussed in Sec. \ref{Sec3}, if the separatrix becomes too
large, one can not distinguish between the captured and the not captured into
resonance trajectories, as the captured trajectories might end up outside
the numerical measurement window. To analyze this effect, one must examine
the size of the separatrix. We begin by writing the Hamiltonian associated
with Eq. (\ref{TiltPend}),
\begin{equation}
H\left( \Phi ,\frac{d\Phi }{d\tau }\right) =\frac{1}{2}\left( \frac{d\Phi }{%
d\tau }\right) ^{2}-4\cos k_{r}P_{1}P_{2}\cos \Phi +\Phi ,  \label{Ham}
\end{equation}%
where the resonance condition (\ref{HamEqsA}) yields $\cos k_{r}=\sqrt{%
1-\left( \frac{\pi \tau }{P_{2}N}\right) ^{2}}$. The separatrix is the
trajectory for which $H$ equals the value of the potential at its maximum
point. Inserting this value of $H$ into (\ref{Ham}) and shifting $\Phi $
such that $\Phi =0$ at the maximum point of the potential, we find the
equation for the separatrix:

\begin{equation}
\left. \frac{d\Phi }{d\tau }\right\vert _{sep}^{\pm }=\pm 2^{1/2}\sqrt{%
B\left( 1-\cos \Phi \right) +\sin \Phi -\Phi },  \label{Sepa}
\end{equation}%
where $B=\sqrt{\left( 4\cos k_{r}P_{1}P_{2}\right) ^{2}-1}$. Following the
arguments in Sec. \ref{Sec3}, we demand that the lower end of the separatrix
in $k,\Phi $ phase-space at the final driving time is higher than the lower
end of our measurement window located at $k=\pi /4$. Thus, we invert Eq. (%
\ref{HamEqsA}) and insert (\ref{Sepa}) to get the condition

\begin{equation}
\arcsin \left[ \left( \left. \frac{d\Phi }{d\tau }\right\vert
_{sep}^{-}+\tau _{f}\right) \frac{\pi }{P_{2}N}\right] >\frac{\pi }{4}.
\label{kSepaCond}
\end{equation}

The dashed-dotted line in Fig. \ref{Fig4} is calculated numerically based on the
limiting case of (\ref{kSepaCond}).



\begin{thebibliography}{99}
\bibitem{SideNote1} When $N\geq 4$ is even, the sum could vanish if $%
l+m^{\prime \prime}=N/2 $ or $l+m^{\prime \prime}=3N/2$. However, in all of these cases, the resulting
couplings link modes below and above $N/2$. Since we use initial
condition $\left\vert a_{0}\right\vert =1$ and the resonant pathway leads
the population to higher modes, one of the modes in the coupling terms
always vanishes, making the couplings irrelevant.
\bibitem{SideNote2} In this context, the nonresonant coupling neglected in
Eq. (\ref{2Level}) is associated with the second energy crossing occurring
at $\tau =-\tau _{c}$, which the driver "misses" by starting at $\tau =0$.
This peculiar effect could be seen numerically by starting the driver at
well before $-\tau _{c}$, yielding a double transition - from mode $0$ to $1$
and back, as the drive sweeps through both energy crossings.
\bibitem{SideNote3} The small excitation below the threshold is the result
of non-resonant and non-adiabatic effects. Unlike other systems \cite%
{Ex5,PhaseSpace}, the finite nature of the energy ladder makes it impossible
to create separation between resonantly and non-resonantly excited
populations.
\end{thebibliography}
\end{document}